\documentclass[twocolumn]{aastex631}

\usepackage{bm}

\usepackage{amsmath}
\usepackage{bbding}
\usepackage{ragged2e}
\usepackage{multirow}

\newcommand{\rev}[1]{#1}  
\newcommand{\reb}[1]{#1}  
\newcommand{\rec}[1]{#1}  
\newcommand{\ree}[1]{#1}

\shorttitle{BADMET}
\shortauthors{Xia et al.}

\begin{document}

\title{Black hole ADaptive Multi-relation Mass Estimation based on Timing Analysis (BADMET)}

\author[0009-0005-3916-1455]{Ruisong Xia\textsuperscript{\Envelope}}
\author[0000-0002-1935-8104]{Yongquan Xue\textsuperscript{\Envelope}}

\affiliation{Department of Astronomy, University of Science and Technology of China, Hefei 230026, China; xiars@ustc.edu.cn, xuey@ustc.edu.cn}
\affiliation{School of Astronomy and Space Science, University of Science and Technology of China, Hefei 230026, China}

\author[0000-0001-5525-0400]{Hao Liu}

\affiliation{Yuxi Normal University, No. 134 Fenghuang Road, Yuxi, Yunnan, 653100, China}

\author[0000-0002-1497-8371]{Guowei Ren}
\affiliation{Department of Astronomy, University of Science and Technology of China, Hefei 230026, China; xiars@ustc.edu.cn, xuey@ustc.edu.cn}
\affiliation{School of Astronomy and Space Science, University of Science and Technology of China, Hefei 230026, China}

\author[0000-0002-0771-2153]{Mouyuan Sun}
\affiliation{Department of Astronomy, Xiamen University, Xiamen, Fujian 361005, China}

\author[0000-0003-4773-4987]{Qingwen Wu}
\affiliation{Department of Astronomy, School of Physics, Huazhong University of Science and Technology, Luoyu Road 1037, Wuhan, China}

\author[0009-0001-9170-3363]{Jialai Wang}

\affiliation{Department of Astronomy, University of Science and Technology of China, Hefei 230026, China; xiars@ustc.edu.cn, xuey@ustc.edu.cn}
\affiliation{School of Astronomy and Space Science, University of Science and Technology of China, Hefei 230026, China}
\affiliation{Kavli Institute for the Physics and Mathematics of the Universe (Kavli IPMU, WPI), The University of Tokyo Institutes for Advanced Study (UTIAS), The University of Tokyo, Kashiwa, Chiba 277-8583, Japan}
\affiliation{Center for Data-Driven Discovery, Kavli IPMU (WPI), UTIAS, The University of Tokyo, Kashiwa, Chiba 277-8583, Japan}

\author[0009-0005-2801-6594]{Shuying Zhou}
\affiliation{Department of Astronomy, Xiamen University, Xiamen, Fujian 361005, China}

\author[0000-0001-8515-7338]{Zhen-Bo Su}
\author[0000-0002-1653-4969]{Shifu Zhu}
\author[0009-0003-5280-0755]{Mengqiu Huang}
\affiliation{Department of Astronomy, University of Science and Technology of China, Hefei 230026, China; xiars@ustc.edu.cn, xuey@ustc.edu.cn}
\affiliation{School of Astronomy and Space Science, University of Science and Technology of China, Hefei 230026, China}

\begin{abstract}
Active galactic nucleus (AGN) variability is widely recognized as a promising avenue for constraining black hole mass, yet its practical application remains limited. 
Although the correlation between black hole mass and variability timescale has been extensively investigated, finite observational baselines often lead to systematic underestimation of the timescale, thereby hindering reliable mass estimation for individual sources. 
To address this limitation, we develop the Black hole ADaptive Multi-relation Mass Estimation based on Timing analysis (BADMET), an extendable framework that provides black hole mass estimates by assimilating multiple mass-related parameters that include variability properties.
We show that incorporating the variability timescale improves mass recovery in the simulated sample and enhances agreement with the reference masses in the observed sample, achieving uncertainties of $\lesssim 0.5$ dex even with a single X-ray observation.


\end{abstract}

\keywords{Active galactic nuclei; Active galaxies; X-ray active galactic nuclei}

\section{INTRODUCTION}\label{sec_intro}

Black hole mass measurements are essential for probing the co-evolution of galaxies and their central supermassive black holes, as well as the history of black hole growth across \rev{cosmic time}.
Multiple methods have been developed to measure the masses of nuclear black holes in galaxies, including both direct and indirect techniques \citep[e.g.,][]{2014SSRv..183..253P}.
Direct measurements, such as stellar or gas dynamics \citep[e.g.,][]{2005SSRv..116..523F} and reverberation mapping \citep{1982ApJ...255..419B}, rely on resolved motions or time-delay measurements and are applicable only to a limited number of nearby or well-monitored sources.
Indirect methods, in contrast, infer black hole masses from scaling relations with other observable properties, including host-galaxy characteristics \citep[e.g.,][]{2000ApJ...539L...9F,2000ApJ...539L..13G,2002ApJ...574..740T}, luminosity \citep[e.g.,][]{1998ApJ...505L..83L, 1999ApJ...526..579W}, or variability \citep[e.g.,][]{2022A&A...666A.127A, 2023MNRAS.521.2954S}, enabling estimates for large samples of \rev{sources}, though with larger uncertainties compared to direct measurements.

Among these indirect approaches, variability from active galactic nuclei (AGNs) has emerged as a particularly promising avenue \citep[e.g.,][]{2014SSRv..183..253P}.
Characteristic timescales \reb{($\tau$), associated with breaks in X-ray or optical} power spectral densities, encode information about accretion processes and correlate with black hole \rev{masses} \citep[e.g.,][]{2004MNRAS.348..207P, 2013ApJ...779..187K, burke_characteristic_2021, 2023MNRAS.521.2954S, 2024ApJ...975..160R, 2026ApJ..1002..201X}. 
Although these relations currently exhibit significant scatters and are limited by observational cadence and baseline \citep[e.g.,][]{2021AcA....71..103K, 2024ApJ...966....8Z}, they provide a scalable method applicable to large samples and offer complementary constraints to traditional scaling relations.
\rec{In our previous work \citep{2025MNRAS.544L..96X}, we develop a baseline-aware framework to account for the systematic underestimation of characteristic timescales caused by finite observational baselines when inferring their dependence on AGN physical properties.
The framework models how an intrinsic timescale is mapped onto the measured timescale given the observational baseline, and incorporates this mapping into the inference within a Bayesian framework.
Compared with conventional selection criteria, such as restricting the analysis to $\tau < 0.1\times{\rm baseline}$ \citep[e.g.,][]{burke_characteristic_2021, 2024ApJ...975..160R}, our method makes use of a broader range of available timescale measurements and provides a nearly unbiased recovery of the underlying correlations.
These results demonstrate the ability of the framework to extract useful physical information from otherwise biased timescale measurements, providing the basis for incorporating the characteristic timescale into black hole mass inference in this work.}

Building on these insights, we introduce a Bayesian framework that unifies multiple observational constraints within a probabilistic formulation, especially dealing with the baseline issues in the scaling relation with variability and black hole mass.
We develop a method, named the Black hole ADaptive Multi-relation Mass Estimation based on Timing analysis (BADMET), which primarily estimates black hole mass from variability information, while remaining extensible to incorporate other measurement techniques.
\rev{BADMET incorporates potentially underestimated variability timescales into mass inference}, integrating available observational parameters (see Section~\ref{sec_method}) to enhance the overall reliability of black hole mass estimates even when individual measurements are highly uncertain (see Section~\ref{sec_perf}), thereby providing a flexible and extensible framework for black hole mass estimation \rev{with} forthcoming time-domain observations (see Sections~\ref{sec_discuss} and~\ref{sec_sum}).

\section{METHODOLOGY}\label{sec_method}

BADMET \ree{(v1.0.0; publicly available on Zenodo; \citealt{xia_badmet_2026})} is developed as a companion algorithm to BADDAT \rev{(i.e., Baseline-Aware Dependence fitting for DAmping Timescales; \citealt{2025MNRAS.544L..96X})}, extending its framework to enable black hole mass inference within the same probabilistic paradigm. 
The derivation proceeds in a similar manner, but with a distinct goal.
The target distribution is now the posterior probability distribution of the black hole mass \rev{for an individual source}, which can be expressed as
\begin{equation}\label{eqo}
P( \log M \,|\, \{Y_{{\rm obs},i}\})
= \frac{P(\{Y_{{\rm obs},i}\} \,|\, \log M) \, P(\log M)}{P(\{Y_{{\rm obs},i}\})},
\end{equation}
where \rec{$M$ is the numerical value of the black hole mass in units of} the solar mass, $i$ indexes the different observational constraints for the source, and the quantity $Y_{{\rm obs},i}$ is obtained from observations, either directly or through an empirical relation.

\reb{Assuming that the constraints are conditionally independent given $\log M$ and adopting a Gaussian likelihood for each constraint, we write}
\begin{equation}\label{eq:like}
    \begin{aligned}
    \ln\mathcal{L}
    &= \ln P(\{Y_{{\rm obs},i}\}\mid \log M)\\
    &= -\frac{1}{2}\sum_i
    \left[
        \frac{(Y_{{\rm m},i}-Y_{{\rm obs},i})^2}{s_i^2}
        +\ln(2\pi s_i^2)
    \right],
    \end{aligned}
\end{equation}
where $Y_{{\rm m},i}$ is the corresponding model value for a given $\log M$ and $s_i^2$ is the total variance for the $i$th constraint.
BADMET accommodates four types of observational constraints detailed as follows, with representative examples listed in Table~\ref{tab1}.

\begin{table*}[t]
\centering
\small
\setlength{\tabcolsep}{4pt}
\caption{Representative examples of the four types of observational constraints supported by BADMET.
}\label{tab1}
\begin{tabular}{clllll}
\hline
\hline
Case & Representative Relation & $Y_{{\rm m},i}$ & $Y_{{\rm obs},i}$ & Additional Inputs & Associated Errors \\
\hline
(1)
& $\log(\tau_{\rm in})=1.28\log M-4.37+\log(1+z)$
& $\log\tau_{\rm m}$
& $\log\tau$
& $z,\ \mathrm{baseline},\ N_{\rm lc}$
& $\sigma_{\epsilon,i},\ \Delta\log\tau_{\rm m},\ \Delta z$ \\

(2)
& $\log(L_{\rm X,m})=0.89\log M+36.72$
& $\log L_{\rm X,m}$
& $\log L_{\rm X}$
& ---
& $\sigma_{\epsilon,i},\ \Delta\log L_{\rm X}$ \\

(3)
& $\log \widehat M=-0.94\log(\sigma_{\rm NXV}^{2}/0.005)+7.41$
& $\log M$
& $\log \widehat M(\sigma_{\rm NXV}^{2})$
& $\sigma_{\rm NXV}^{2}$
& $\sigma_{\epsilon,i},\ \Delta\sigma_{\rm NXV}^{2}$ \\

(4)
& External mass estimate
& $\log M$
& $\log M_{\rm est}$
& ---
& $\Delta \log M_{\rm est}$ \\
\hline
\end{tabular}
\justifying
\tablecomments{
\reb{The representative relations for Cases~(1)--(3) are adopted from \citet{2026ApJ..1002..201X}, \citet{2009A&A...501..915B}, and \citet{2022A&A...666A.127A}, respectively.
For Case~(1), the relation describes the timescale before underestimation, while the predicted output $\log\tau_{\rm m}$ is obtained through Equation~\eqref{eq:tau_model}.
Additional inputs include auxiliary quantities and observing conditions.
The listed errors are combined according to Equations~\eqref{eq:var_tau}, \eqref{eq:var_forward}, \eqref{eq:var_inverse}, and \eqref{eq:var_direct} for Cases~(1)--(4), respectively.
\rec{For compactness, the numerical values of timescales and X-ray luminosities are understood to be in units of seconds and $\mathrm{erg\,s^{-1}}$, respectively.}
}}
\end{table*}

{
\textbf{(1) Variability timescale relations.}
When the observational baseline is insufficiently long relative to the intrinsic variability timescale, \rec{damped random walk (DRW)} fitting can systematically underestimate the timescale \citep[e.g.,][]{2009ApJ...698..895K, 2017A&A...597A.128K, burke_characteristic_2021, 2024ApJ...961....5H, 2024ApJ...975..160R, 2024ApJ...966....8Z}.
We account for this bias by modeling the expected measured timescale for a given intrinsic timescale \citep{2025MNRAS.544L..96X, 2026ApJ..1002..201X}, resulting in a probability distribution $P(\log \rho_{\rm out}|\log \rho_{\rm in})$, which captures the underestimation caused by the limited observational baseline. Here, $\rho$ is defined as the timescale normalized by the observational baseline, i.e., $\rho=\tau/\mathrm{baseline}$. \rec{Here and throughout this work, the numerical values of timescales and baselines are taken in seconds when used in expressions.}

We take $Y_{{\rm obs},i}=\log(\tau)$, where $\tau$ is the observer-frame timescale estimated by fitting a DRW model to the light curve via maximum likelihood \citep{celerite}.
The corresponding model value is \rec{$\log \tau_{\rm m}$},
\begin{equation}\label{eq:tau_model}
   Y_{{\rm m},i}
    =\log \tau_{\rm m}
    =\xi(\mu)+\log\left(\mathrm{baseline}\right),
\end{equation}
where $\xi(\mu)$ is the conditional expectation of
$\log\rho_{\rm out}$ given $\log\rho_{\rm in}=\mu$,
as determined by the conditional probability distribution
$P(\log\rho_{\rm out}\mid\log\rho_{\rm in})$.
This distribution captures the timescale underestimation caused by the finite baseline. It depends on both the observational baseline and the number of light-curve data points, $N_{\rm lc}$, with further details \rev{elucidated} in \citet{2025MNRAS.544L..96X}.

Here, $\mu$ represents the modeled logarithm of the intrinsic observer-frame timescale normalized by the observational baseline, i.e., $\log \rho_{\rm in,m}$, before accounting for the underestimation due to limited baseline, written as
\begin{equation}\label{eq:tau_input}
    \begin{aligned}
    \mu={}&k_0 \log M+b+f(\{X_j\})-\log(\mathrm{baseline})
    \end{aligned},
\end{equation}
where $k_0$ is the coefficient of $\log M$, $b$ is the intercept, and $f(\{X_j\})$ describes the contribution of the auxiliary quantities, with $j$ indexing these quantities.
For example, when redshift is the only auxiliary quantity, $\{X_j\}=\{z\}$ and $f(\{X_j\})=\log(1+z)$.
The baseline is the time span of the light curve in the observer frame.

The total variance is
\begin{equation}\label{eq:var_tau}
    s_i^2=
    \sigma_{\epsilon,i}^2
    +[\Delta\log\tau_{\rm m}]^2
    +\sum_j
    \left(
        \frac{\partial\log\tau_{\rm m}}{\partial X_j}\Delta X_j
    \right)^2.
\end{equation}
Here, the quantity $\sigma_\epsilon$ denotes the scatter of the adopted timescale relation, and \rec{since $\log \tau_m$ and $\log \rho_m$ differ only by the logarithm of the baseline of the specific light curve, their uncertainties are identical:} $[\Delta\log\tau_{\rm m}]^2=[\Delta\log\rho_{\rm m}]^2=\operatorname{Var}(\log\rho_{\rm out}\mid\log\rho_{\rm in}=\mu)$.
The final term propagates the measurement errors in the auxiliary quantities ($\Delta X_j$), including the redshift, through the timescale relation and the mapping $\xi(\mu)$.

\textbf{(2) Forward scaling relations.}
A forward relation predicts an observational quantity from $\log M$ and the auxiliary quantities:
\begin{equation}\label{eq:forward_model}
    Y_{{\rm m},i}=f_i(\log M,\{X_j\}),
\end{equation}
which is compared with its observed counterpart $Y_{{\rm obs},i}$.
For example, these quantities can be the predicted and observed logarithmic X-ray luminosities.
The total variance is
\begin{equation}\label{eq:var_forward}
    s_i^2=
    \sigma_{\epsilon,i}^2
    +[\Delta Y_{{\rm obs},i}]^2
    +\sum_j
    \left(
        \frac{\partial Y_{{\rm m},i}}{\partial X_j}\Delta X_j
    \right)^2.
\end{equation}
Here, $\sigma_{\epsilon,i}$ denotes the scatter of the adopted relation.
The remaining terms account for the observational uncertainty in $Y_{{\rm obs},i}$ and the propagated measurement errors in the auxiliary quantities.

\textbf{(3) Inverse mass-scaling relations.}
An inverse relation provides a mass estimate $\log \widehat M(\{X_j\})$ from observed quantities.
This estimate is compared directly with the mass parameter $\log M$ being inferred:
\begin{equation}\label{eq:inverse_model}
    Y_{{\rm obs},i}=\log \widehat M(\{X_j\}),
    \qquad
    Y_{{\rm m},i}=\log M.
\end{equation}
The total variance is
\begin{equation}\label{eq:var_inverse}
    s_i^2=
    \sigma_{\epsilon,i}^2
    +\sum_j
    \left(
        \frac{\partial \log\widehat M}{\partial X_j}\Delta X_j
    \right)^2.
\end{equation}
Here, $\sigma_{\epsilon,i}$ denotes the scatter of the adopted mass-scaling relation.
The sum propagates the measurement errors in the input quantities into the derived mass estimate.
The hat distinguishes the derived estimate $\log \widehat M$ from the parameter $\log M$ being inferred.

\textbf{(4) Independent mass estimates.}
An external mass estimate can be incorporated directly as
\begin{equation}\label{eq:direct_model}
    Y_{{\rm obs},i}=\log M_{\rm est},
    \qquad
    Y_{{\rm m},i}=\log M,
\end{equation}
with
\begin{equation}\label{eq:var_direct}
    s_i^2=(\Delta \log M_{\rm est})^2.
\end{equation}
In this case, BADMET directly adopts the externally reported mass estimate $\log M_{\rm est}$ and its uncertainty $\Delta \log M_{\rm est}$, without requiring the underlying measurements or reconstructing the uncertainty budget.}

Following the user guide of \textsc{EMCEE} \citep{foreman-mackey_emcee_2013a} with the Equations~\eqref{eqo}~and~\eqref{eq:like}, the data can be well modeled to derive the black hole mass for each source.

\section{PERFORMANCE}\label{sec_perf}
\subsection{Recovery Tests on Simulated Data}\label{sec_rec}

To assess the reliability of BADMET, we perform a recovery test using a mock dataset.
We generate a mock sample containing \ree{1000} sources with black hole masses ($\log M_{\rm true}$) and corresponding X-ray observational properties, including the light curve baseline, the number of data points ($N_{\rm lc}$) of the light curve, and the redshift ($z$).
The mock sample is generated by drawing parameters from uniform distributions within the following ranges: 
$\log M_{\rm true} \in [5, 9]$, baseline $\in [1, 14]\times10^{4}~\mathrm{s}$, 
$N_{\rm lc} \in [50, 1400]$, and $z \in [0, 0.3]$.
\rec{These ranges are chosen to fully encompass the corresponding parameter ranges of the real reference sample described in Section~\ref{sec_real_test}.}
We simulate three representative measurement cases, as illustrated by examples of cases (1), (2), and (3) in Table~\ref{tab1}.

The first \rev{case} is the timescale measurement, and its relation with black hole mass is \citep{2026ApJ..1002..201X}:
\begin{equation}\label{eq_tau}
    \log \tau_{\rm in} = (1.28\pm0.10)\log M - (4.37\pm0.66) + \log(1 + z).
\end{equation}
The mock observed timescales $\tau$ are then generated by sampling from the conditional probability distribution $P(\log \tau_{\rm out}|\log \tau_{\rm in})$ \citep{2025MNRAS.544L..96X}, which accounts for the effects of the \rec{limited} baseline and $N_{\rm lc}$ of the light curves.
The scatter of $\tau$ in this relation is simulated as 0.4 \rev{dex}, estimated by \citet{2026ApJ..1002..201X}.

The second \rev{case} is the \rec{absorption-corrected} 2--10 keV luminosity ($L_{\rm X}$) measurement, which is considered to correlate with black hole mass \citep{2009A&A...501..915B}:
\begin{equation}\label{eq_LX}
    \log  L_{\mathrm{X}}  = (0.89\pm0.07) \log M + (36.72\pm 0.06),
\end{equation}
\rev{where $L_{\rm X}$ is the numerical value of X-ray luminosity in units of $\mathrm{erg\,s^{-1}}$.}
The scatter of this relation is set to $\sigma_{\epsilon}=0.7$, a typical value for this relation.
The mock observational uncertainty for the luminosity is $\Delta(\log L_{\rm X})=0.2$.

The \rev{third case} is the normalized excess variance ($\sigma_{\rm NXV}^2$) measurement, defined as \citep{1997ApJ...476...70N}:
\begin{equation}\label{eq_NXV_def}
    \sigma^2_{\rm NXV}=\frac{1}{N_{\rm lc}\bar{R}^2}\sum_{i=1}^{N_{\rm lc}}\big((R_i-\bar{R})^2-\sigma_i^2\big),
\end{equation}
where $R_i$ and $\sigma_i$ are the count rates and their uncertainties, respectively, and $\bar{R}$ is the mean count rate.
The corresponding black hole mass relation is given by \citet{2022A&A...666A.127A}:
\begin{equation}\label{eq_NXV}
    \log M(\sigma_{\mathrm{NXV}}^{2}) = -(0.94\pm0.06) \log ( \frac{\sigma_{\mathrm{NXV}}^{2}}{0.005} ) + (7.41\pm0.05),
\end{equation}
with a scatter of $0.7$ dex. The observational uncertainty \rev{of $\log \sigma_{\mathrm{NXV}}^{2}$ is assumed to be 0.02 dex, approximately the median value obtained from the error propagation in our previous work \citep{2026ApJ..1002..201X}.}
\ree{We adopt this relation as a representative example, although our sample does not fully match its calibration conditions and we do not impose all of its applicability criteria.
This adopted relation serves as an approximate example and can be updated as improved calibrations become available.}

When generating the mock data, the uncertainties of all parameters $\Delta Y_{{\rm m}}$ and $\Delta X_j$, and the coefficients of the equations $\Delta \frac{\partial Y_{{\rm m},i}}{\partial X_j}$ are considered.
The observational uncertainty $\Delta Y_{{\rm obs}, i}$ is also simulated except for $\tau$.
\reb{As the fluctuations of $\tau$ in the maximum-likelihood estimate are already represented by the conditional distribution, no additional independent measurement-error term is introduced.}

\begin{figure*}[!]
    \centering
    \includegraphics[width=1\linewidth]{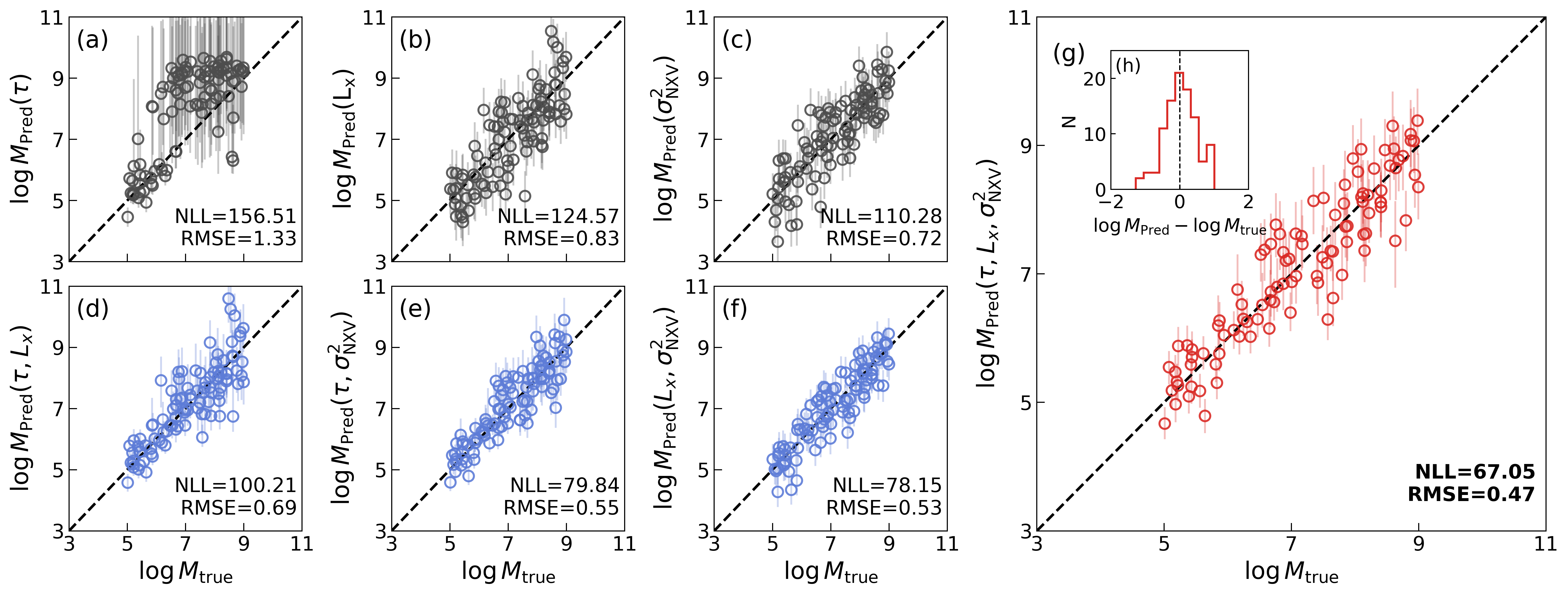}
    \caption{Comparison of the true black hole masses with those recovered from seven different models \ree{for 100 sources randomly selected from} the mock sample, encompassing all combinations of the three mock parameters ($\tau$, $L_{\rm X}$, $\sigma^2_{\rm NXV}$\rev{; see Table \ref{tab1}}).
    Error bars indicate the 16th and 84th percentiles of the posterior distributions obtained from Markov Chain Monte Carlo sampling. 
    The diagonal dashed line represents the one-to-one relation. For each panel, the NLL and the RMSE are shown in the lower right corner to quantify the recovery accuracy, with the NLL in the combined model highlighted in bold as the lowest, emphasizing its superior performance among the models.
    \rev{The inset panel} (h) shows the histogram of the differences between predicted and \rev{true} black hole masses, $\log M_{\rm pred} - \log M_{\rm true}$.
    }
    \label{fig1}
\end{figure*}

For each mock source, we estimate the black hole mass using seven combinations of the three \rec{aforementioned} cases.
Figure~\ref{fig1} compares the true and predicted black hole masses for \ree{100 sources} randomly selected from the mock \ree{sample, for clarity,} under each combination. 
\ree{The results for the full sample of 1000 sources are reported in the ``Correct $\sigma_\epsilon$’’ column of Table~\ref{tabB}.}
The recovery performance is quantified using the negative log-likelihood (NLL) and the root-mean-square error (RMSE), which reflect the reliability and uncertainty consistency of the measurements and the overall accuracy of the recovered masses, respectively.
Panel~(a) shows that using the timescale alone yields relatively accurate mass estimates at the low-mass end, but the estimates become poorly constrained at the high-mass end, resulting in the largest overall RMSE of \ree{1.33~dex}.
This is expected because the limited observational baseline leads to underestimation of the longer variability timescales associated with higher-mass black holes \citep[e.g.,][]{2025MNRAS.544L..96X}.
Panels~(b) and~(c) show that $L_{\rm X}$ and $\sigma_{\rm NXV}^2$ alone recover the overall mass trend, with RMSEs of \ree{0.83 and 0.72~dex}, respectively.
Incorporating the timescale improves both estimates, reducing the RMSE to 0.69~dex for the combination of $\tau$ and $L_{\rm X}$ in panel~(d), and to 0.55~dex for $\tau$ and $\sigma_{\rm NXV}^2$ in panel~(e).
Combining $L_{\rm X}$ and $\sigma_{\rm NXV}^2$ also yields an RMSE of \ree{0.53~dex}, as shown in panel~(f).
Finally, combining all three parameters in panel~(g) gives the lowest NLL and RMSE of \ree{67.05 and 0.47~dex}, respectively, with the residuals concentrated near zero \rec{shown} in the inset.
These results indicate that, although the model based solely on X-ray luminosity or $\sigma^2_{\rm NXV}$ provides a reasonable estimate, incorporating variability information in the combined model improves the recovery, producing tighter agreement with the true masses. In particular, a comparison between panels (f) and (g) shows that adding the characteristic timescale reduces the RMSE from \ree{0.53 to 0.47 dex}.

\subsection{Consistency Tests on Observational Data}\label{sec_real_test}

\ree{
To evaluate the practical applicability of BADMET, we apply it to the X-ray sample of \citet{2026ApJ..1002..201X}, comprising 112 sources and 399 exposures.
For each exposure, we consider three mass indicators: the characteristic timescale $\tau$, the X-ray luminosity $L_{\rm X}$, and the normalized excess variance $\sigma_{\rm NXV}^{2}$. 
The values of $\tau$ and $L_{\rm X}$ are taken directly from our previous work \citep{2026ApJ..1002..201X}, while $\sigma_{\rm NXV}^{2}$ is calculated from the X-ray light curves with Equation~\eqref{eq_NXV_def}.}

\begin{figure*}[!]
    \centering
    \includegraphics[width=\linewidth]{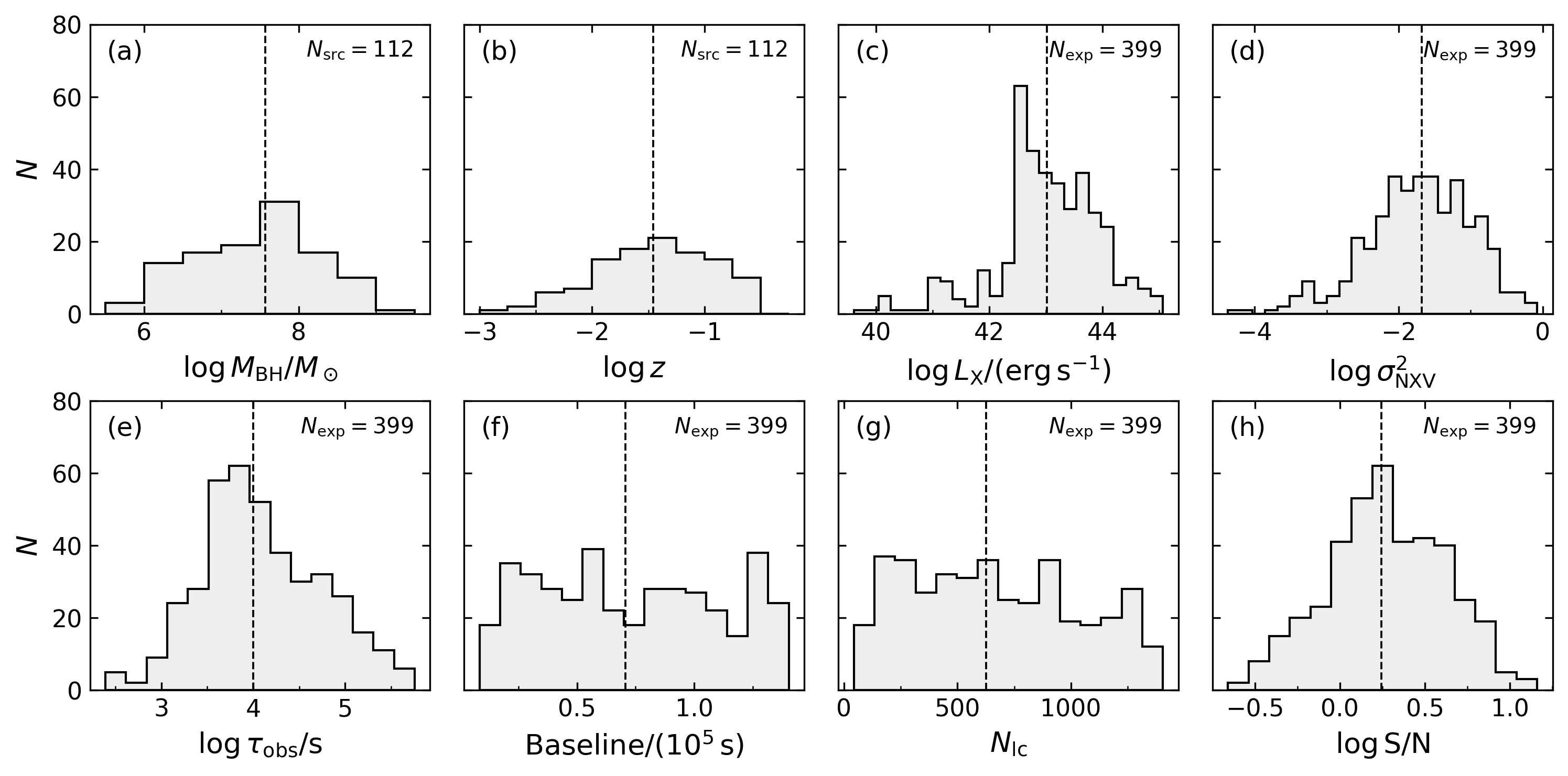}
    \caption{
    \ree{
    Distributions of the properties of the real X-ray test sample:
    (a) black hole mass, (b) redshift, (c) X-ray luminosity,
    (d) normalized excess variance, (e) observer-frame DRW timescale,
    (f) observing baseline, (g) number of light-curve data points, and
    (h) DRW signal-to-noise ratio.
    Panels (a) and (b) are evaluated for the 112 distinct sources, while
    panels (c)--(h) are evaluated for the 399 individual exposures.
    Dashed lines mark the median values, and the annotations give the
    number of measurements included in each distribution.
    }}
    \label{fig2}
\end{figure*}

\ree{The sample is selected from XMM-Newton EPIC-pn observations in the 4XMM-DR14 catalogue, with an initial requirement of a $2.0\text{--}4.5$ keV flux above $10^{-13}\ {\rm erg\,cm^{-2}\,s^{-1}}$ and an exposure time longer than 10 ks. 
The light curves are extracted in the $0.3\text{–}10$ keV energy band with a time bin size of 100 s.
To obtain reliable variability measurements, \citet{2026ApJ..1002..201X} require significant variability relative to white noise and apply quality criteria to the DRW fits based on the Akaike information criterion. 
The final sample includes only sources with literature black hole mass estimates.
The distributions of the sample properties and observational characteristics are shown in Figure~\ref{fig2}. 
Black hole mass and redshift are counted once per source, whereas $\tau$, $\sigma_{\rm NXV}^{2}$, $L_{\rm X}$, the number of light-curve data points $N_{\rm lc}$, the baseline, and the DRW signal-to-noise ratio (S/N, defined as the ratio of the DRW amplitude to the noise) are shown for individual exposures.
}

Each exposure is analyzed independently, resulting in a single mass estimate per exposure (see more \rev{discussions} in Appendix~\ref{secA}).
Figure~\ref{fig3} presents the comparison between the predicted black hole masses and the literature values for the real X-ray sample.
Similar to the mock tests, including \ree{timescale} information in the combined model effectively improves the mass estimates, reducing uncertainties and achieving better agreement with the literature values compared to models using only timescale, X-ray luminosity, or \rec{$\sigma_{\rm NXV}^2$.
When} considering all three parameters together, the RMSE of the mass measurement is reduced to 0.45\rec{~dex} in this sample.
The median posterior uncertainty, defined as half the interval between the 16th and 84th percentiles, is 0.43~dex, comparable to the RMSE.
Note that if the scatters of the relations are unknown, the measurements may be moderately affected (see Appendix~\ref{secB} for details).

\section{DISCUSSIONS}\label{sec_discuss}

\subsection{The Role of Variability Timescales}
\begin{figure*}[!]
    \centering
    \includegraphics[width=1\linewidth]{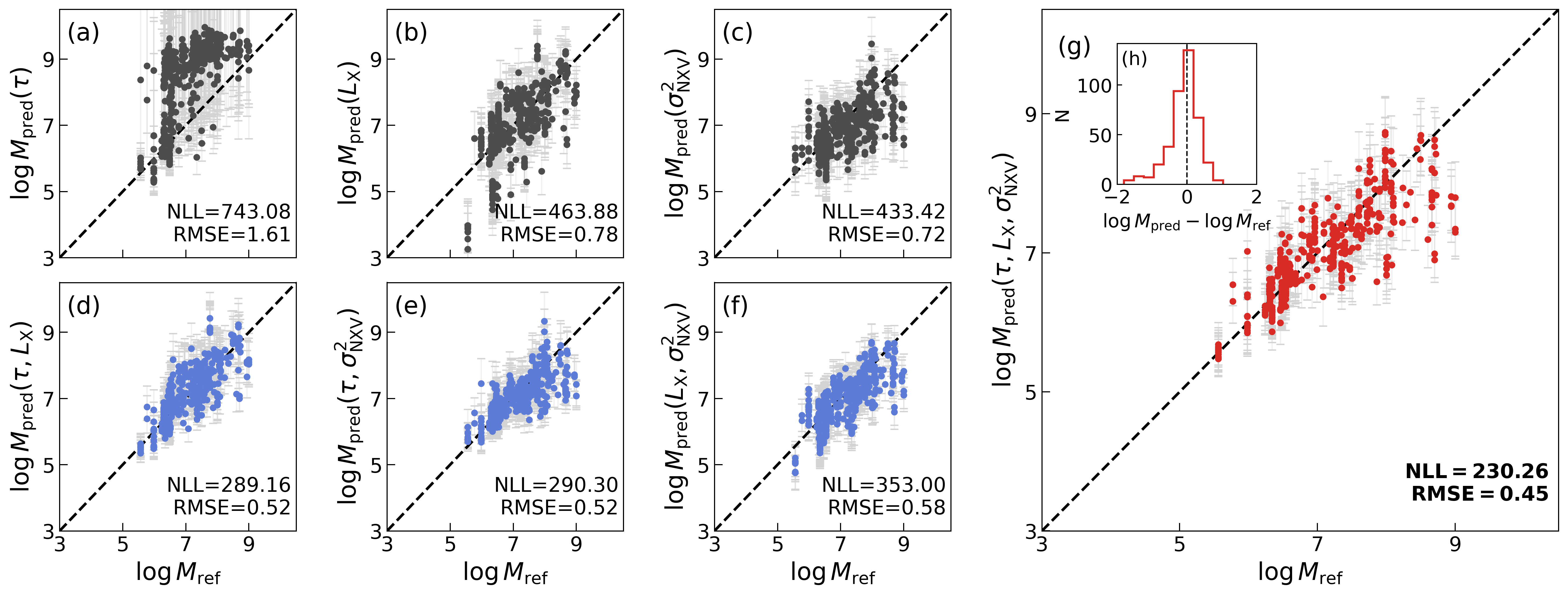}
    \caption{\rev{Same as} Figure~\ref{fig1} but for the real X-ray sample \citep{2026ApJ..1002..201X}. \rec{The reference black hole masses are adopted from the literature, with a typical uncertainty of 0.1 dex; see Table A1 of \citet{2026ApJ..1002..201X} for details.}}
    \label{fig3}
\end{figure*}

BADMET inherits the idea of BADDAT while being further developed specifically for inferring black hole masses. 
As shown in the panels of $\log M_{\rm pred}(\tau)$ in Figures~\ref{fig1} and~\ref{fig3}, estimating black hole \rev{masses} solely from variability timescales is unreliable for a large proportion of the data, as these observational baselines are generally insufficient to capture robust timescales. 
However, when combining $\tau$ with other mass-related estimations, the inference of black hole mass is significantly improved. 
The contribution of $\tau$ is evident in both mock and observational samples. 

This is reasonable because for lower-mass black holes, $\tau$ can be measured more accurately and reduces the scatter of estimations, while for higher-mass black holes, $\tau$ \reb{can provide a probabilistic lower constraint} of black hole mass that compensates for potential underestimation by other parameters.
Specifically, some black hole masses appear severely underestimated when using $L_{\rm X}$ alone (see Figure~\ref{fig3} in this work and also Figure~3 of \citealt{2009A&A...501..915B}), and combining the two measurements of $L_{\rm X}$ and $\tau$ effectively resolves these discrepancies, yielding more accurate mass estimates.

\subsection{Assumptions and Applicability of BADMET}

BADMET assumes that the target sources follow the adopted black hole mass scaling relations and that the adopted scatters adequately describe the source population. 
We adopt a uniform prior on $\log(M_{\rm BH}/M_\odot)$ over [0,12], sufficiently broad to encompass possible black hole masses while minimizing the influence of prior boundaries.
When timescale information is included, we assume that the adopted conditional probability model $P(\log\rho_{\rm out}\mid\log\rho_{\rm in})$ adequately describes the effects of finite observational baselines and sampling on the inferred timescales.
We further assume that the mass indicators are conditionally independent given the black hole mass, allowing their individual likelihoods to be multiplied.

The empirical relations are not built into BADMET but should be selected by users, who are encouraged to adopt the most recent and robustly established relations available.
While BADMET is designed as a broadly extensible framework, its applicability in any specific black hole mass estimation depends on the domains covered by the adopted empirical relations.
Each relation should include a scatter term, $\sigma_{\epsilon,i}$, to quantify the tightness of its correlation with black hole mass. 
The input quantities should have definitions, units, and energy bands compatible with those used to establish the corresponding scaling relations, or be appropriately converted.
Measurement uncertainties and relation scatters must be reasonably characterized because they determine the relative weight and concentration of each likelihood contribution, affecting both the inferred mass and its posterior uncertainty (see more details in Appendix~\ref{secB}).

For relations involving variability timescales, the correlations should be established using timescales measured through DRW fitting through maximum likelihood estimation (MLE), consistently with the present framework, and the light curve must be adequately fitted by a DRW model.
Other timescale definitions, such as the bending timescale of a bending power-law power spectral density (PSD), should not be directly substituted because their correspondence depends on the underlying PSD shape.

Some empirical relations include additional parameters to better describe the variation in an observable, but these parameters do not necessarily improve the constraints on black hole mass. 
Simpler relations may therefore be preferable when the additional parameters provide little information for mass inference.
For example, in the best-fitting model of \citet{2026ApJ..1002..201X}, $\tau$ depends on both $\log M$ and the X-ray photon index $\Gamma$. 
Although $\tau$ and $\Gamma$ are significantly correlated, $\Gamma$ is only weakly correlated with $\log M$ (Pearson correlation coefficient $r=-0.19$). 
We test the relation including $\Gamma$ and find little change in the inferred masses (not shown here for conciseness), and therefore adopt the relation without $\Gamma$ in Case~(1).
Nevertheless, parameters required by the adopted empirical relation or providing additional information for mass inference should be included as auxiliary variables $X_j$, such as redshift when using the variability timescale relation.

The assumption of conditional independence in Equation~\eqref{eq:like} neglects potential correlations among mass indicators at fixed black hole mass. As detailed in Appendix~\ref{secB} and Table~\ref{tabB}, such correlations can affect both the accuracy of the inferred masses and the calibration of their posterior uncertainties, with the magnitude and direction of the effects depending on the indicators and their correlations. 
In most of the tested cases, the RMSE differs by less than approximately 0.1 dex from the value obtained for the same indicator combination without residual correlations. Although these dependencies could be modeled using a covariance matrix, the required information is generally unavailable in practical applications. We therefore retain the conditional-independence assumption as an approximation, while recognizing that the resulting posterior uncertainties may be either underestimated or overestimated.

\subsection{Limitations of the Current Validation}

The present observational validation is restricted to relatively X-ray-bright AGNs with significant variability, acceptable DRW fits, and available literature black hole mass estimates. 
Its performance under less favorable observational conditions remains to be assessed.
In the observational tests, literature masses serve as reference estimates and carry their own statistical and systematic uncertainties, limiting a more in-depth test of the capability of BADMET.
Regarding the choice of empirical relations, the relations for $L_{\rm X}$ and $\sigma_{\rm NXV}^{2}$ are derived from independent samples, whereas the timescale relation is derived from the same sample used in the present observational test, limiting the independence of this validation.
This limitation is difficult to avoid because, to our knowledge, no well-established X-ray DRW timescale–black hole mass relation derived from an independent sample is currently available.

The histogram in panel (h) of Figure~\ref{fig3} appears to be skewed toward negative values for the observational sample. 
Several possible explanations can be identified from panels (a)–(c). 
The most direct indication is seen in panel (c), where the black hole masses inferred from $\sigma_{\rm NXV}^{2}$ tend to be underestimated at the high-mass end, corresponding to a larger \(\sigma_{\rm NXV}^{2}\) than expected from the adopted scaling relation. 
This may partly reflect limitations in the empirical relation. 
The relations presented by \citet{2022A&A...666A.127A} rely on several assumptions and are subject to specific applicability conditions, while a more generally applicable empirical relation has yet to be established.
In particular, the effects of light-curve sampling on variability amplitude measurements have not been systematically characterized and warrant further investigation.
Another possible explanation is a selection effect. 
Since the variability amplitudes tend to decrease with increasing black hole mass, detecting significant variability becomes more challenging at the high-mass end. 
Consequently, sources that pass the variability selection criteria and can be successfully fitted with the DRW model may be preferentially biased toward larger variability amplitudes, leading to underestimated black hole masses.
Additional contributions may arise from the relatively large scatter in the mass estimates based on $L_{\rm X}$ and the weak constraints provided by $\tau$ at the high-mass end, as indicated by the large uncertainties in panel (a) and the scatter in panel (b).

Despite these limitations, the agreement between our inferred masses and the literature reference values supports the applicability of BADMET to the sample studied here.

\section{\ree{Summary and Future Prospects}}\label{sec_sum}

\ree{
This study provides a proof of concept for incorporating potentially underestimated variability timescales into joint black hole mass inference, making use of the complementary information provided by different parameters derived from a single X-ray observation.
For data of sufficient quality and under the adopted scaling relations, our mock and observational tests demonstrate that black hole masses can be constrained to a typical uncertainty of approximately 0.5 dex using a single X-ray observation.
}

The ongoing XMM-Newton \citep{2001A&A...365L...1J} and \rev{Einstein Probe} \citep{2022hxga.book...86Y} surveys are expected to deliver extensive X-ray variability data, allowing for new and more accurate time-domain black hole mass measurements across AGN samples \rev{\citep[e.g., ][]{2018SSPMA..48c9508X}}.
\ree{Furthermore, BADMET can incorporate mass-related information from any measurement, provided that a reliable scaling relation with black hole mass is available. 
This flexibility allows the framework to be extended beyond X-ray data to optical and other wavelength bands, as well as to incorporate newly established scaling relations, potentially enabling more accurate black hole mass estimates.}

\begingroup
\let\internallinenumbers\relax
\let\endinternallinenumbers\relax

\begin{acknowledgments}
The research has made use of data obtained from the 4XMM XMM-Newton Serendipitous Source Catalog compiled by the 10 institutes of the XMM-Newton Survey Science Centre selected by ESA. This work is supported by the National Key R\&D Program of China (2023YFA1608100), NSFC grants (12025303 and 12393814). 
G.W.R acknowledges support from the Anhui Provincial Natural Science Foundation  (2508085QA007) and the China Postdoctoral Science Foundation (grant No. 2025M773191). S.Y.Z. acknowledges support from the National Natural Science Foundation of China (NSFC-125B2058).
\end{acknowledgments}
\endgroup

\bibliography{sample631}{}

\appendix
\renewcommand{\thetable}{B\arabic{table}}
\setcounter{table}{0}
\renewcommand\thefigure{\Alph{section}\arabic{figure}} 
\setcounter{figure}{0}

\section{Multiple Exposures of a Single Source}\label{secA}

\begin{figure*}[h]
    \centering
    \includegraphics[width=1\linewidth]{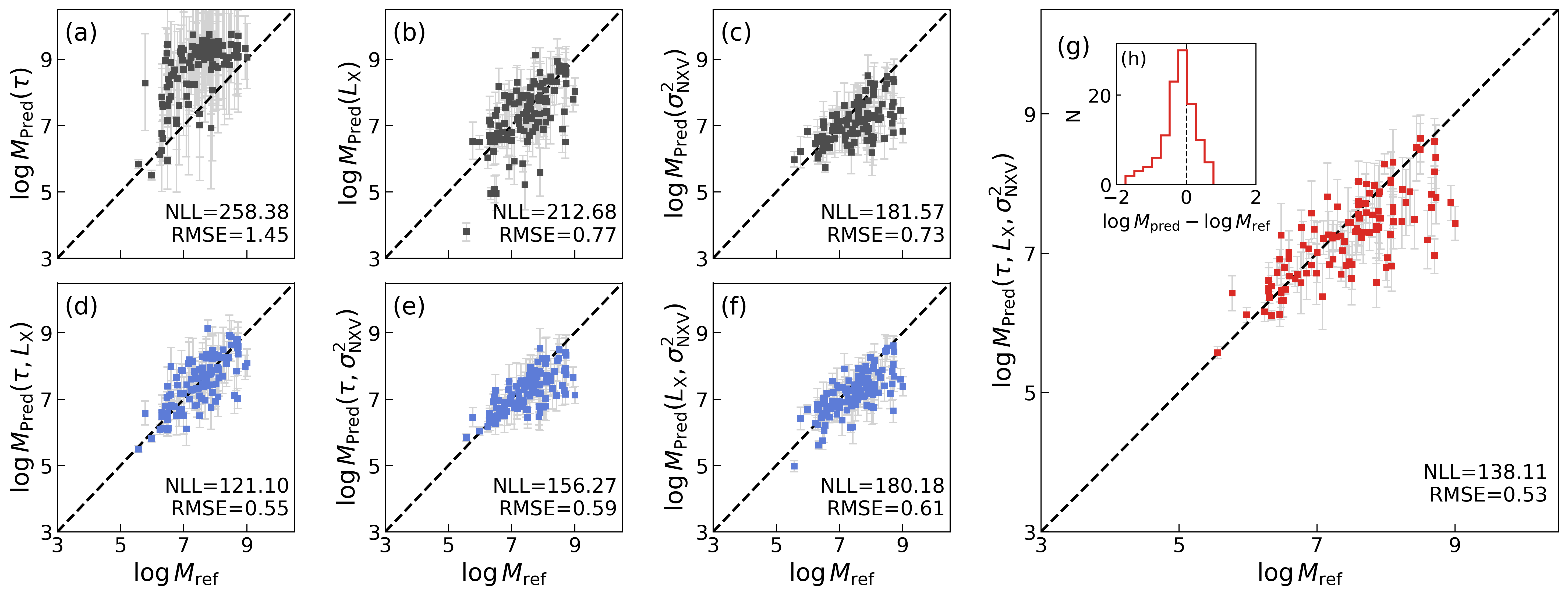}
    \caption{\rev{Same as} Figure~\ref{fig3}, but showing the final mass estimates for sources. \rec{For sources with multiple exposures, the information from those exposures is combined.}}
    \phantomsection
    \label{figA}
\end{figure*}

Multiple exposures of a single source can be regarded as multiple estimations of a given black hole mass, collectively contributing to the final mass estimate of the source.
These estimations can be combined by BADMET once the errors are well estimated, corresponding to Case (4) in Table~\ref{tab1}.
Figure~\ref{figA} presents the results for each source \rev{(i.e., only one final data point for one source)}. 
While the NLL for the combination of the three parameters is no longer the smallest, the RMSE remains the lowest (see panel~(g)), indicating that although the estimated uncertainties may not fully capture the true scatter, the resulting mass estimates still show the best agreement with the reference masses in this sample.
The RMSE for the entire sample increases compared to Figure~\ref{fig3}, primarily because sources with multiple accurately measured exposures are treated as single data points, affecting the overall error statistics. 
Nevertheless, the RMSE remains at an acceptable level of $\sim0.5$~dex.

\setcounter{figure}{0}

\section{\ree{Sensitivity to the Assumed Relation Scatter and Residual Covariance}}
\label{secB}

\ree{
Since the relations between the parameters and the black hole mass are empirical and typically exhibit scatter, BADMET includes an additional error term $\sigma_{\epsilon,i}$.
This scatter term depends on both the sample properties and observational conditions.
To quantify the impact of incorrectly specifying the relation scatters, we repeat the mock-sample analysis assuming several common values of \(\sigma_\epsilon\), using the correctly specified, parameter-specific relation scatters as the reference case. 
The relation-scatter assumption section of Table~\ref{tabB} reports the RMSE of the recovered black hole masses and the fraction of true masses enclosed by the posterior 16th--84th percentile intervals (\(C_{68}\)).
All metrics are calculated using the sample of 1000 mock sources described in Section~\ref{sec_rec}.
}

\ree{
We see that using the correct relation scatters produces well-calibrated uncertainties, with $C_{68}=0.65$--$0.68$ across the tested sets, consistent with the nominal coverage of 0.68.
Adopting an obviously low common scatter of $0.1$~dex leads to substantial undercoverage.
An assumed scatter of $0.6$~dex, representing a slight deviation from the preset scatters, generally yields coverage closer to the nominal 68\% level, whereas $0.9$~dex, representing a substantial overestimation, leads to systematic overcoverage.
The RMSE is less sensitive to the assumed scatter than the interval coverage.
For example, for the estimate considering all three parameters, the RMSE changes from $0.44$~dex with the correct indicator-specific scatters to $0.56$~dex when a common scatter of $0.1$~dex is assumed.
The accuracy seems relatively insensitive to the assumed relation scatters, except when the scatter is substantially underestimated, whereas the reliability of the inferred uncertainties depends more strongly on the assumed scatters.
}

\ree{
BADMET currently combines the likelihoods under the assumption that their intrinsic residuals are conditionally independent at fixed black hole mass. 
However, since these indicators are derived from the same observation and may be intrinsically related even at fixed black hole mass, the residuals of their empirical relations may be correlated.
The observable-residual covariance section of Table~\ref{tabB} explores the effect of violating the conditional-independence assumption by introducing pairwise correlations between the relation residuals. 
The impact depends on both the correlated pair and the indicator combination used. 
Relative to the correctly specified reference results in the relation-scatter section of Table~\ref{tabB}, the largest absolute changes are approximately 0.18~dex in RMSE and 0.24 in \(C_{68}\).
Both occur for the two-parameter \(L_{\rm X}\) and \(\sigma^2_{\rm NXV}\) combination at \(\rho=0.6\).
In this case, the RMSE decreases from 0.53 to 0.35~dex, likely because the correlated residuals partially cancel when propagated through relations with opposite slopes, whereas \(C_{68}\) increases from 0.66 to 0.90, indicating overly conservative uncertainty intervals.
For most other configurations, the absolute changes in both RMSE and \(C_{68}\) remain below approximately 0.10, although the coverage can shift to either side of 0.68, resulting in overconfident or conservative uncertainty estimates.
Given that the residual covariance is generally unknown a priori and has a relatively modest impact in most configurations, we retain the conditional-independence assumption as a practical approximation, while recognizing it as a limitation of the current analysis.
}

\begin{table*}[t]
\centering
\caption{\ree{Mock-sample tests of relation-scatter and covariance assumptions.}}
\phantomsection
\label{tabB}
\small
\renewcommand{\arraystretch}{1.08}
\setlength{\tabcolsep}{3pt}
\begin{tabular*}{\textwidth}{@{\extracolsep{\fill}}lrrrrrrrrrr@{}}
\multicolumn{11}{l}{\textbf{Relation-scatter assumption}} \\
\hline
Indicators & \multicolumn{2}{c}{Correct $\sigma_\epsilon$} & \multicolumn{2}{c}{$\sigma_\epsilon=0.1$} & \multicolumn{2}{c}{$\sigma_\epsilon=0.3$} & \multicolumn{2}{c}{$\sigma_\epsilon=0.6$} & \multicolumn{2}{c}{$\sigma_\epsilon=0.9$} \\
 & RMSE & $C_{68}$ & RMSE & $C_{68}$ & RMSE & $C_{68}$ & RMSE & $C_{68}$ & RMSE & $C_{68}$ \\
\hline
$\tau$ & 1.40 & 0.68 & 1.41 & 0.48 & 1.40 & 0.62 & 1.40 & 0.77 & 1.36 & 0.87 \\
$L_{\rm X}$ & 0.81 & 0.68 & 0.81 & 0.24 & 0.81 & 0.38 & 0.81 & 0.62 & 0.81 & 0.81 \\
$\sigma^2_{\rm NXV}$ & 0.72 & 0.65 & 0.71 & 0.11 & 0.71 & 0.32 & 0.71 & 0.60 & 0.72 & 0.79 \\
$\tau, L_{\rm X}$ & 0.63 & 0.68 & 0.67 & 0.25 & 0.66 & 0.42 & 0.65 & 0.69 & 0.66 & 0.84 \\
$\tau, \sigma^2_{\rm NXV}$ & 0.56 & 0.65 & 0.63 & 0.12 & 0.59 & 0.37 & 0.58 & 0.65 & 0.58 & 0.83 \\
$L_{\rm X}, \sigma^2_{\rm NXV}$ & 0.53 & 0.66 & 0.62 & 0.12 & 0.54 & 0.34 & 0.53 & 0.60 & 0.53 & 0.80 \\
$\tau, L_{\rm X}, \sigma^2_{\rm NXV}$ & 0.44 & 0.65 & 0.56 & 0.12 & 0.46 & 0.35 & 0.46 & 0.64 & 0.46 & 0.83 \\
\hline
\multicolumn{11}{l}{\textbf{Residual covariance}} \\
\hline
Indicators & \multicolumn{2}{c}{Correlated pair} & \multicolumn{2}{c}{$\rho=-0.6$} & \multicolumn{2}{c}{$\rho=-0.3$} & \multicolumn{2}{c}{$\rho=0.3$} & \multicolumn{2}{c}{$\rho=0.6$} \\
 & \multicolumn{2}{c}{} & RMSE & $C_{68}$ & RMSE & $C_{68}$ & RMSE & $C_{68}$ & RMSE & $C_{68}$ \\
\hline
$\tau, L_{\rm X}$ & \multicolumn{2}{c}{$\tau$--$L_{\rm X}$} & 0.58 & 0.75 & 0.61 & 0.70 & 0.69 & 0.62 & 0.73 & 0.59 \\
$\tau, \sigma^2_{\rm NXV}$ & \multicolumn{2}{c}{$\tau$--$\sigma^2_{\rm NXV}$} & 0.62 & 0.62 & 0.59 & 0.63 & 0.52 & 0.73 & 0.50 & 0.77 \\
$L_{\rm X}, \sigma^2_{\rm NXV}$ & \multicolumn{2}{c}{$L_{\rm X}$--$\sigma^2_{\rm NXV}$} & 0.67 & 0.57 & 0.60 & 0.62 & 0.44 & 0.78 & 0.35 & 0.90 \\
$\tau, L_{\rm X}, \sigma^2_{\rm NXV}$ & \multicolumn{2}{c}{$\tau$--$L_{\rm X}$} & 0.41 & 0.72 & 0.43 & 0.70 & 0.46 & 0.65 & 0.48 & 0.62 \\
$\tau, L_{\rm X}, \sigma^2_{\rm NXV}$ & \multicolumn{2}{c}{$\tau$--$\sigma^2_{\rm NXV}$} & 0.49 & 0.62 & 0.47 & 0.65 & 0.42 & 0.71 & 0.40 & 0.75 \\
$\tau, L_{\rm X}, \sigma^2_{\rm NXV}$ & \multicolumn{2}{c}{$L_{\rm X}$--$\sigma^2_{\rm NXV}$} & 0.55 & 0.60 & 0.50 & 0.64 & 0.38 & 0.74 & 0.32 & 0.83 \\
\hline
\end{tabular*}
\justifying
\tablecomments{
Both tests use a mock sample containing 1000 sources. Correct $\sigma_\epsilon$ denotes the parameter-specific generating relation scatters. $\rho$ is the correlation between observable-space relation residuals.
}
\end{table*}

\end{document}